\documentclass[preprint,authoryear,5p,twocolumn]{elsarticle}
\usepackage{graphicx,epsfig}
\usepackage{dcolumn}
\usepackage{bm}
\usepackage{psfrag}
\usepackage{plain}
\usepackage{subfig}
\usepackage{tabularx}
\usepackage[latin1]{inputenc}
\usepackage{amsmath}
\usepackage{amssymb}
\usepackage{graphics}
\usepackage{graphicx}
\usepackage{epsfig}

\usepackage{amssymb}
\usepackage{amsthm}

\journal{Geomorphology}

\begin{document}

\begin{frontmatter}


\title{Model for a dune field with exposed water table}


\author[ufc]{Marco C. M. de M. Luna\corref{mcmml}}\ead{marcocesarluna@gmail.com}\author[ufc,deq]{Eric J. R. Parteli}\author[ufc,eth]{Hans J. Herrmann}
\address[ufc]{Departamento de F\'{\i}sica, Universidade Federal do Cear\'a - 60455-760, Fortaleza, CE, Brasil.}
\address[deq]{Programa de P\'os-Gradua\c{c}\~ao em Engenharia Qu\'{\i}mica, Universidade Federal do Cear\'a, 60455-900, Fortaleza, CE, Brazil.}
\address[eth]{Institut f\"ur Baustoffe IfB, ETH H\"onggerberg, HIF E 12, CH-8093, Z\"urich, Switzerland.}


\begin{abstract}
Aeolian transport in coastal areas can be significantly affected by the presence of an exposed water table. In some dune fields, such as in Len\c{c}\'ois Maranhenses, northeastern Brazil, the water table oscillates in response to seasonal changes of rainfall and rates of evapotranspiration, rising above the ground during the wet season and sinking below in the dry period. A quantitative understanding of dune mobility in an environment with varying groundwater level is essential for coastal management as well as for the study of long-term evolution of many dune fields. Here we apply a model for aeolian dunes to study the genesis of coastal dune fields in presence of an oscillating water table. We find that the morphology of the field depends on the time cycle, $T_{\mathrm{w}}$, of the water table and the maximum height, $H_{\mathrm{w}}$, of its oscillation. Our calculations show that long chains of barchanoids alternating with interdune ponds such as found at Len\c{c}\'ois Maranhenses arise when $T_{\mathrm{w}}$ is of the order of the dune turnover time, whereas $H_{\mathrm{w}}$ dictates the growth rate of dune height with distance downwind. We reproduce quantitatively the morphology and size of dunes at Len\c{c}\'ois Maranhenses, as well as the total relative area between dunes.
\end{abstract}


\end{frontmatter}




\section{\label{sec:introduction}Introduction}

It is widely accepted that dune morphology depends fundamentally on the wind directionality and on the amount of sand available for transport \citep{Wasson_and_Hyde_1983}. While longitudinal and star dunes form under bi- and multidirectional wind regimes, respectively, the best understood types of dune are formed by unidirectional wind. In this case, crescent-shaped barchans occur if the sand availability is low, while transverse dunes, which display nearly invariant profile in the direction orthogonal to the wind, form if the ground is covered with sand. The shape of dunes can be significantly modified due to natural agents such as vegetation growth \citep{Hesp_2002,Tsoar_and_Blumberg_2002,Barbosa_and_Dominguez_2004} or cementation of sand by mineral salts \citep{Schatz_et_al_2006}. Modeling has brought many insights on dune formation in environments with stabilizing vegetation \citep{Nishimori_and_Tanaka_2001,Baas_2002,Duran_and_Herrmann_2006a,Yizhaq_et_al_2007,Nield_and_Baas_2008a,Nield_and_Baas_2008b,Luna_et_al_2011}. The influence of sand induration on the shape of dunes has been also investigated \citep{Herrmann_et_al_2008,Rubin_and_Hesp_2009}. 

The water table can also play a fundamental role in sediment transport on coasts and in deserts \citep{Kocurek_et_al_1992,Ruz_and_Meur_Ferec_2004,Kocurek_et_al_2007,Mountney_and_Russell_2009}. The dynamics of the water table in some aeolian sand systems is connected with seasonal variations in climate and rainfall \citep{Levin_et_al_2009}, and has been modeled by many different authors \citep{de_Castro_Ochoa_and_Munoz_Reinoso_1997,Kocurek_et_al_2001}. However, quantitatively little is known about dune field morphodynamics in presence of a varying groundwater level. 

\begin{figure*}[!ht]
\begin{center}
{\includegraphics[width=0.93\textwidth]{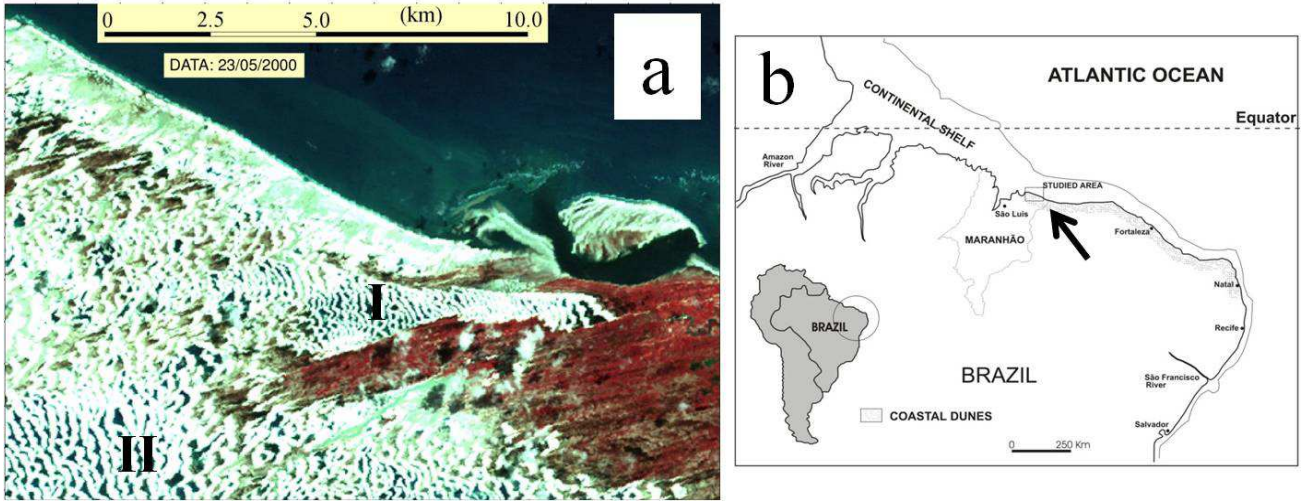}}
\caption{(a) Coastal dune field named ``Len\c{c}\'ois Maranhenses'' in the State of Maranh\~ao, northeastern Brazil, located near $2^{\circ}35^{\prime}$S, $42^{\circ}49^{\prime}$W (Landsat image of May 23th of 2000). North is to the top. The landscape of the field is dominated by chains of barchanoids, alternating with water ponds, as those in areas I and II. (b) The arrow in the map indicates the location of the field.}
\label{fig:satelite_image_and_map}
\end{center}
\end{figure*}

One example of dunes evolving in presence of a dynamic water table is the coastal dune field known as ``Len\c{c}\'ois Maranhenses'', located in the State of Maranh\~ao, northeastern Brazil (Fig.~\ref{fig:satelite_image_and_map}). Long chains of laterally linked barchans (the so-called ``barchanoids'') extending over several kilometers constitute the dominant dune morphology at Len\c{c}\'ois (Fig.~\ref{fig:satelite_image_and_map}a). Dunes at Len\c{c}\'ois alternate with freshwater lagoons that form in the rainy seasons and, at some places, nearly disappear in the summer. In spite of the field research undertaken in the past \citep{Goncalves_et_al_2003,Parteli_et_al_2006,Levin_et_al_2007,Kadau_et_al_2009}, it has remained unclear how the oscillating water table is contributing to the long-term dynamics and to the shape of dunes at Len\c{c}\'ois. 

Understanding the effect of the water table on dune field evolution may help predict coastal dune mobility and elucidate the history of many ancient deserts \citep{Davidson_Arnott_and_Pyskir_1988,Enzel_et_al_1999,Mountney_and_Thompson_2002,Chen_et_al_2004,Grotzinger_et_al_2005,Bourke_and_Wray_2011}. A physically based model that accounts for a mathematical description of sand transport and for the evolution of the terrain in response to aeolian drag and variations in the groundwater level is required. 

In the present work, we adapt a recently developed continuum model for aeolian dunes \citep{Sauermann_et_al_2001,Kroy_et_al_2002} to study the formation of coastal dune fields in an area of exposed groundwater. This model has been applied to study different types of dune encountered in nature, and has proven to reproduce the shape of real dunes yielding quantitative agreement with measurements \citep{Sauermann_et_al_2003,Parteli_et_al_2006,Duran_and_Herrmann_2006a,Duran_and_Herrmann_2006b,Parteli_and_Herrmann_2007,Duran_et_al_2008,Parteli_et_al_2009,Parteli_et_al_2011}. The dune model encodes a mathematical description of the turbulent wind field over the terrain with a continuum model for saltation --- which consists of grains hopping in ballistic trajectories close to the ground and ejecting new particles upon collision with the bed \citep{Bagnold_1941}. Here we add to the dune model a water table that can rise and sink seasonally, thus affecting local rates of sand transport during the evolution of the dune field. In particular, we aim to understand the genesis and evolution of dunes in presence of a dynamic water level in order to shed light on the conditions leading to the dune morphologies observed at Len\c{c}\'ois Maranhenses.

In the next Section, we present a brief introduction about climate and wind regime of the area of Len\c{c}\'ois Maranhenses. In Section \ref{sec:model} we present a description of the dune model and its extension in order to account for the water table. The calculations of coastal dune fields are described in Section \ref{sec:simulations}, while in Section \ref{sec:results} we present and discuss our results. Conclusions are presented in Section \ref{sec:conclusions}.


\section{\label{sec:lencois}Len\c{c}\'ois Maranhenses}

The National Park of Len\c{c}\'ois Maranhenses is located on the coastal area of the Maranh\~ao State, northeastern Brazil. The area comprises 155 thousand hectares delimited by the coordinates S $02^{\circ}19^{\prime}$ and $02^{\circ}45^{\prime}$, and W $42^{\circ}44^{\prime}$ and $43^{\circ}29^{\prime}$ (Figs. 1a,b). Climate at Len\c{c}\'ois is semi-humid tropical with sparse vegetation, air humidity of 68$\%$ and annual average temperature about $28.5^{\circ}$C \citep{IBAMA_2003,Floriani_et_al_2004}. 

\begin{figure}[!t]
\begin{center}
{\includegraphics[width=1.0\columnwidth]{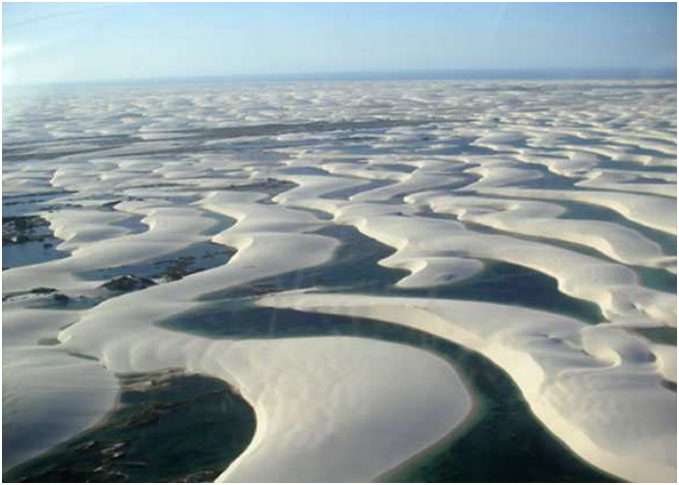}}
\caption{Aerial photo of barchanoids and interdune ponds at Len\c{c}\'ois Maranhenses.}
\label{fig:barchanoids_aerial_photo}
\end{center}
\end{figure}

The landscape of Len\c{c}\'ois is characterized by the presence of long chains of barchanoids (Fig.~\ref{fig:barchanoids_aerial_photo}) giving the impression of a crumpled sheet --- hence the origin of the name: ``Len\c{c}\'ois'' means ``sheets'' in portugese. These dunes extend along 75 km and penetrate inland to distances larger than 20 km. They are detached from the coast by a deflation plane of width between 600 m and 2000 m, and are migrating on top of dunes of older generations \citep{Tsoar_et_al_2009}. Insights on the formative process and early development stages of dunes at Len\c{c}\'ois Maranhenses have been gained from recent field research \citep{Goncalves_et_al_2003,Parteli_et_al_2006}. Sand deposited by tides on the beach is eroded by the wind thus forming small barchans with height between 50~cm and 1~m close to the beach \citep{Parteli_et_al_2006}. As the barchans advance downwind, they become larger, link laterally and give place to barchanoidal chains that can reach heights of 30~m. The sand of the dunes is composed by quartz grains of mean diameter varying between $120 {\mu}$m and $350 {\mu}$m \citep{IBAMA_2003}, which are values around the average diameter $d = 250 \mu$m previously reported for grains of sand dunes in other fields \citep{Bagnold_1941,Pye_and_Tsoar_1990}.

\begin{figure}[!ht]
\begin{center}
{\includegraphics[width=0.77\columnwidth]{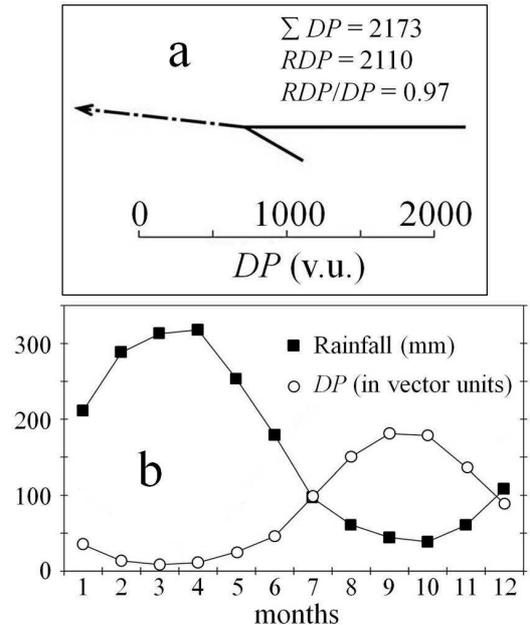}}
\caption{(a) Annual sand rose of Arana\'u (for 2003), on the coast of western Cear\'a, located at $2^{\circ}49^{\prime}$S, $40^{\circ}13^{\prime}$W. The sand rose shows that most sand-moving winds come from the east. The continuous lines indicate the yearly total wind power ({\it{DP}}) from the corresponding directions, while the dashed-dotted arrow shows the downwind direction of the resultant drift potential ({\it{RDP}}). The high value of {\it{RDP}}$/${\it{DP}} is consistent with a strongly unimodal wind regime. (b) Rainfall and wind power, monthly averaged for the years 1991-2007. The data were taken from the NCEP/NCAR Reanalysis \citep{Tsoar_et_al_2009}. Wind power and rainfall data are for $2.5^{\circ}$S, $40^{\circ}$W, and $2.9^{\circ}$S, $39.4^{\circ}$W, respectively.}
\label{fig:sand_rose_and_rainfall}
\end{center}
\end{figure}

Wind regime at Len\c{c}\'ois is strongly unidirectional. The corresponding sand rose shows that the strongest winds, which can reach velocities over 8~m/s \citep{Jimenez_et_al_1999}, blow mainly from the East (c.f. Fig.~\ref{fig:sand_rose_and_rainfall}a). The vector length of each direction of the sand rose gives the potential rate of sand transport from that direction, or the drift potential \citep{Tsoar_2001},
\begin{equation}
DP = \sum k\,u^2{\left[{u-u_{\mathrm{t}}}\right]}f,
\end{equation}
where $k \approx 7.3$, $u$ is the wind velocity (in m/s) at a height of 10~m, $u_{\mathrm{t}} \approx 6.19$~m/s is the threshold wind velocity and $f$ is the fraction of time the wind was above $u_{\mathrm{t}}$ \citep{Tsoar_2001}. The resultant drift potential ({\it{RDP}}) is obtained by calculating the vector sum of the drift potential for each one of the directions of the sand rose. Wind directionality can be then quantified in terms of the ratio $\beta =$ {\it{RDP}}$/${\it{DP}}, where {\it{DP}} is the sum of the magnitude of the drift potential for all directions. A value of $\beta$ close to unity or close to zero means unidirectional or multidirectional wind regime, respectively. At Len\c{c}\'ois, $\beta$ is approximately $0.97$, which is consistent with the unidirectional wind regime of the area \citep{Tsoar_et_al_2009}.

Wind power at Len\c{c}\'ois is negatively correlated with rainfall, as depicted in Fig.~\ref{fig:sand_rose_and_rainfall}b. During the dry season from August to December there is almost no rainfall and month-averaged values of wind velocity are much higher than during the wet season. From January to July, when almost $93\%$ of the rainfall takes place \citep{Jimenez_et_al_1999}, the interdune areas are inundated due to the high water table thus forming ponds (Fig.~{\ref{fig:barchanoids_aerial_photo}). The lagoons, placed amidst very clean sand, have no inlet or outlet and are exclusively filled by rain water. Their bottom is covered by a soft brown or green sheet of algae and cyanobacteria \citep{Kadau_et_al_2009}. The interdune ponds cover $41\%$ of the area of Len\c{c}\'ois. The average interdune pond area is 7~ha, and most of the ponds are longer in the cross-wind direction than in the along-wind direction. Using the definition of \cite{Robinson_and_Friedman_2002}, the ponds at Len\c{c}\'ois have a mean circularity --- where the circularity of a sphere is one --- of about $0.47$ \citep{Levin_et_al_2007}. A large fraction of the interdune lakes formed in the rainy season disappear in the dry season. The landscape of Len\c{c}\'ois appears to change continuously, indeed dune mobility implies that the lakes often reappear in different places with different contours. The effect of the fluctuating water level on dune morphology at Len\c{c}\'ois is still poorly understood.


\section{\label{sec:model}The model}

The model used in the calculations of the present work consists of a set of mathematical equations that describe the average surface shear stress ($\tau(x,y)$) over the topography, the mass flux ($q(x,y)$) of saltating particles and the time evolution of the surface resulting from particle transport \citep{Sauermann_et_al_2001,Kroy_et_al_2002,Duran_and_Herrmann_2006b,Duran_et_al_2010}. Here, the model is extended in order to account for the water table. In the model the following steps are solved in an iterative manner.

(i) Given an initial topography, e.g. a terrain with dunes or smooth sand hills, the average shear stress field over the surface is calculated solving a set of analytical equations developed by \cite{Weng_et_al_1991}. The wind model computes, first, the Fourier-transformed components of the topographically induced perturbations in the average shear stress:
\begin{equation}
{\tilde{{{\hat{\tau}}}}}_x = {\frac{{{\tilde{h}}_{\mathrm{s}}}k_x^2}{|\vec{k}|}}{\frac{2}{U^2(l)}}{\left\{{-1 + {\left({2{\ln{{\frac{l}{z_0}}}} + {\frac
{{|{\vec{k}}|}^2}{k_x^2}}}\right)}{\sigma}{\frac{K_1(2{\sigma})}{K_0(2{\sigma})}}}\right\}},
\end{equation}
\begin{equation}
{\tilde{{{\hat{\tau}}}}}_y = {\frac{{{\tilde{h}}_{\mathrm{s}}}k_xk_y}{|\vec{k}|}}{\frac{2}{U^2(l)}}2{\sqrt{2}}{\sigma}K_1(2{\sqrt{2}}{\sigma}),
\end{equation}
where $x$ and $y$ are the components parallel, respectively, perpendicular to the wind direction, $\vec{k}$ is the wave vector, and $k_x$ and $k_y$ its coordinates in Fourier space; ${\tilde{h}}_{\mathrm{s}}$ is the Fourier transform of the height profile, which is defined as the envelope comprising the sand landscape and the level of exposed water; $U$ is the normalized vertical velocity profile, $l$ is the inner layer depth of the flow and $L$ is $1/4$ the mean wavelength of the Fourier representation of the height profile; $\sigma = \sqrt{{\mbox{i}}Lk_xz_0/l}$, $K_0$ and $K_1$ are modified Bessel functions, and $z_0$ is the aerodynamic roughness of the surface. The average shear stress is obtained, then, from the equation,
\begin{equation}
{\vec{\tau}} = {\tau}_0{({{{\vec{\tau}}_0/{\tau}_0 + {\vec{{\hat{\tau}}}}}})},
\end{equation} 
where ${\vec{\tau}}_0$ is the undisturbed shear stress over the flat ground and ${\tau_0} \equiv |{\vec{\tau}}_0|$.
\begin{figure}[!ht]
\begin{center}
{\includegraphics[width=0.93 \columnwidth]{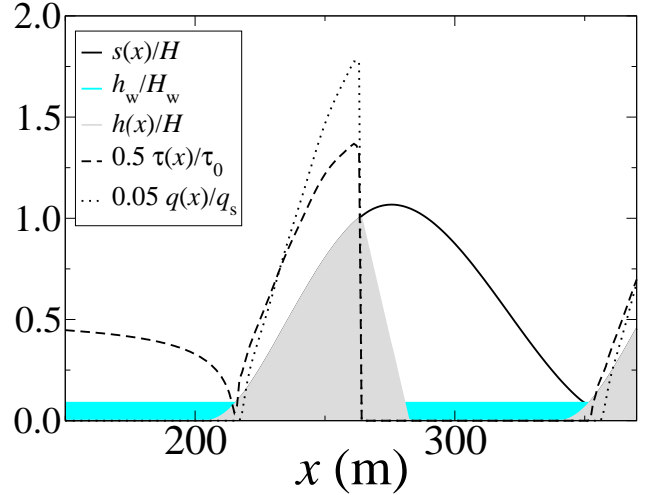}}
\caption{Longitudinal profile of a transverse dune, $h(x)$, and the separation bubble, $s(x)$, both rescaled by the dune height, $H$. The shear stress, ${\tau}(x)$, and the sand flux, $q(x)$, are normalized by the upwind shear stress, ${\tau}_0$, and the saturated flux, $q_{\mathrm{s}}$, respectively. The water level $h_{\mathrm{w}}(x)$ is rescaled by its maximum value, $H_{\mathrm{w}}$. Wind direction is from left to right.}
\label{fig:watertablemodel}
\end{center}
\end{figure} 

The wind model by \cite{Weng_et_al_1991} is valid only for smooth surfaces, and thus the calculation of the wind flow over dunes must be adapted in order to account for flow separation at the dune brink \citep{Kroy_et_al_2002}. For each longitudinal slice of the dune, a separation streamline, $s(x,y)$, is introduced at the dune lee, as depicted in Fig.~\ref{fig:watertablemodel}. The wind model is solved, then, for the envelope, 
\begin{equation}
h_s(x,y) = {\mbox{max}}{\{{h(x,y),s(x,y),h_{\mathrm{w}}}\}}, \label{eq:separation_bubble}
\end{equation} 
comprising the dune surface, $h(x,y)$, the separation streamlines at the dune lee, $s(x,y)$, and the water table, $h_{\mathrm{w}}$. The streamlines of flow separation define the so-called separation bubble, inside which the wind shear is set as zero \cite{Kroy_et_al_2002}. The shape of $s(x,y)$ is approximated by a third-order polynomial, the coefficients of which are calculated from the continuity of $h$, $s$ and their respective first derivatives at the brink and at the reattachment point downwind, which is computed assuming that $s(x,y)$ has a maximum slope \citep{Kroy_et_al_2002}. \newline

(ii) Next, the local sand flux over the landscape above the water level is calculated. The cloud of saltating grains is considered as a thin fluid-like layer moving over the surface. Once the wind speed exceeds the saltation threshold, the sand flux increases, first, exponentially due to multiplicative process inherent to the splash events at grain-bed collisions. However, the wind strength decreases as more grains enter saltation, since these have to be accelerated at cost of aeolian momentum \citep{Nishimura_and_Hunt_2000,Almeida_et_al_2006}. After a saturation distance, the wind is just strong enough to sustain transport and the sand flux is maximal. By using mass and momentum conservation, and by explicitly accounting for the flux saturation transients, the following equation is derived for the height-integrated mass flux of sand per unit time and length \citep{Sauermann_et_al_2001},
\begin{equation}
{\vec{\nabla}}{\cdot}{\vec{q}}  = {\left({1 - {|{\vec{q}}|}/{q_{\mathrm{s}}}}\right)}|{{\vec{q}}}|/{{\ell}_{\mathrm{s}}}, \label{eq:differential_q}
\end{equation}
where $q_{\mathrm{s}} = {({2{v_{\mathrm{s}}{\alpha}}}/{g})}{\rho}_{\mathrm{a}}{u_{{\ast}{\mathrm{t}}}^2}{[{{\left({{u_{\ast}}/u_{{\ast}{\mathrm{t}}}}\right)}^2 - 1}]}$ is the saturated flux and ${\ell}_{\mathrm{s}} = ({2v_{\mathrm{s}}^2{\alpha}/g{\gamma}})/[{{\left({{u_{\ast}}/u_{{\ast}{\mathrm{t}}}}\right)}^2 - 1}]$ the characteristic length of flux saturation, whereas $u_{\ast} = {\sqrt{{\tau}/{\rho}_{\mathrm{a}}}}$ is the wind shear velocity and ${\rho}_{\mathrm{a}}=1.225$ kg$/$m$^3$ is the air density; $u_{{\ast}{\mathrm{t}}}$, the impact threshold \citep{Bagnold_1941}, is about $80\%$ the minimal threshold velocity ($u_{{\ast}{\mathrm{ft}}} \sim 0.26$ m$/$s) required to initiate saltation; $g = 9.81$ m$/$s$^2$ is gravity, and the average grain velocity, $v_{\mathrm{s}}$, is computed by taking the steady-state wind velocity within the saltation layer \citep{Sauermann_et_al_2001,Duran_and_Herrmann_2006b}, whereas $\alpha=0.43$ and $\gamma=0.2$ are empirically determined parameters \citep{Sauermann_et_al_2001,Duran_and_Herrmann_2006b}. 

(iii) The local height, $h(x,y)$, of the sand landscape evolves according to the equation 
\begin{equation}
{\partial}h/{\partial}t = -{\vec{\nabla}}{\cdot}{\vec{q}}/{\rho}_{\mathrm{b}}, \label{eq:mass_conservation}
\end{equation}
where ${\rho}_{\mathrm{b}} = 1650$ kg$/$m$^3$ is the bulk density of the sand. If the local slope exceeds $34^{\circ}$, the unstable surface relaxes through avalanches in the direction of the steepest descent. Avalanches are considered to be instantaneous since their time-scale is much smaller than the one of dune motion. The downslope flux of avalanches is calculated with the equation,
\begin{equation}
{\vec{q}}_{\mathrm{aval}} = k{\left[{ {\mbox{tanh}}({{\nabla}h}) - {\mbox{tanh}}({\theta}_{\mathrm{dyn}}) }\right]}{\frac{{\nabla}h}{|{\nabla}h|}}, \label{eq:avalanche_flux}
\end{equation}
where $k = 0.9$ and ${\theta}_{\mathrm{dyn}} = 33^{\circ}$ is the so-called ``dynamic'' angle of repose \citep{Duran_et_al_2010}. The calculation of the flux due to avalanches followed by the update of the local height through solving Eq.~(\ref{eq:mass_conservation}) is performed iteratively until the local slope is below ${\theta}_{\mathrm{dyn}}$. The calculation of the avalanches applies to the whole sand topography, i.e. including the surface below the water level. The only difference between the transport under and above the water table concerns erosion due to the action of the wind. The surface below the water level is obviously protected from erosion and can only evolve in time due to avalanches following the deposition of sand incoming from dunes above the water level. At the interface with the exposed water, the sand flux vanishes and the incoming volume of sand is, then, instantaneously accreted to the local surface, at the first grid element where the sand surface is below the water level.

(iv) Thereafter, the level of exposed water, $h_{\mathrm{w}}(t)$, is updated. It is assumed that the water table is roughly proportional to rainfall \citep{Tsoar_et_al_2009}, as observed in real situations \citep{Kocurek_et_al_1992}. The water table oscillates in time ($t$) with a period $T_{\mathrm{w}}$ and maximum level $H_{\mathrm{w}}$ according to the equation:
\begin{equation}
h_{\mathrm{w}}(t) = H_{\mathrm{w}}{\mbox{sin}}[{2{\pi}t/T_{\mathrm{w}}}], \label{eq:water_table}
\end{equation}
which is chosen here to represent the nearly sinusoidal behaviour of average rainfall, as monitored over several seasons, in areas of coastal dunes \citep{Jimenez_et_al_1999,Tsoar_et_al_2009} (see Fig.~\ref{fig:sand_rose_and_rainfall}b). 

The calculations are performed using open and periodic boundaries in the directions longitudinal ($x$) and perpendicular ($y$) to the wind, respectively. 

\section{\label{sec:simulations}Simulations of coastal dune fields}

The initial surface is a sand beach, which is modeled as a flat transverse sand hill of height $1.5$ m, width $80$ m and Gaussian longitudinal profile, as depicted in Fig.~\ref{fig:initialsurface}. The transverse profile of the hill has small, random fluctuations of amplitude of the order of the grain diameter, $d \approx 250\,{\mu}$m \citep{Luna_et_al_2011}. The sand hill is subjected to a wind of constant average shear velocity $u_{\ast}$. We take $u_{\ast}$ values within the range $0.3-0.4$ m$/$s, which is typical for sand-moving winds at Len\c{c}\'ois Maranhenses \citep{Jimenez_et_al_1999,Parteli_et_al_2006,Tsoar_et_al_2009}.
\begin{figure}[!ht]
\begin{center}
{\includegraphics[width=1.0\columnwidth]{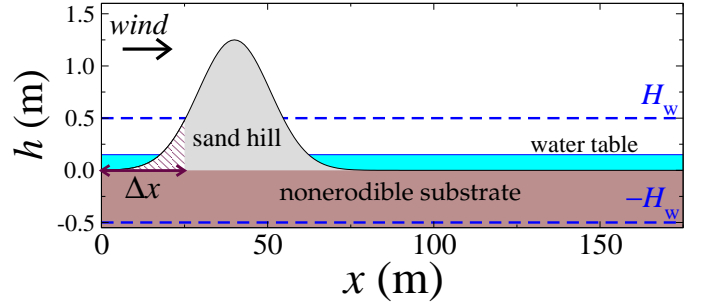}}
\caption{Schematic diagram showing the longitudinal profile of the initial transverse sand hill used in the simulation. The upwind hatched area of the sand hill represents the nonerodible backshore, which has width ${\Delta}x \approx 25$~m and constant profile during the calculation. The maximum and mininum levels of the water table, both indicated by the dashed horizontal lines, are $H_{\mathrm{w}}$ and $-H_{\mathrm{w}}$, respectively.}
\label{fig:initialsurface}
\end{center}
\end{figure} 

The calculation of a coastal dune field in the absence of water table has been studied in previous modeling \citep{Duran_et_al_2010,Luna_et_al_2011}. As shown from these works, the condition for the genesis of a coastal dune field is a saturated influx coming from the sea. In that case net deposition occurs at the windward foot of the sand surface. The hill does not evolve into a migrating dune \citep{Katsuki_et_al_2005}, instead it remains fixed, increases in size and flattens \citep{Duran_et_al_2010}. The flat sand surface resulting from the initial hill is unstable and develops undulations --- the so-called ``sand-wave instabilities'' \citep{Elbelrhiti_et_al_2005} --- which evolve into small transverse dunes of height between 50~cm and 1~m. Once the transverse dunes reach the bedrock, they become unstable and decay into barchans \citep{Reffet_et_al_2010,Parteli_et_al_2011}. Average dune size increases with distance as small barchans merge and collide with larger, more slowly migrating dunes downwind. In this manner, a sand hill subjected to a saturated flux becomes a source of sand for a field of barchans. Figure \ref{fig:nowater} shows a snapshot of the calculation of dune field genesis in the absence of water table \citep{Duran_et_al_2010,Luna_et_al_2011}. 
\begin{figure*}[!ht]
\begin{center}
{\includegraphics[width=0.8 \textwidth]{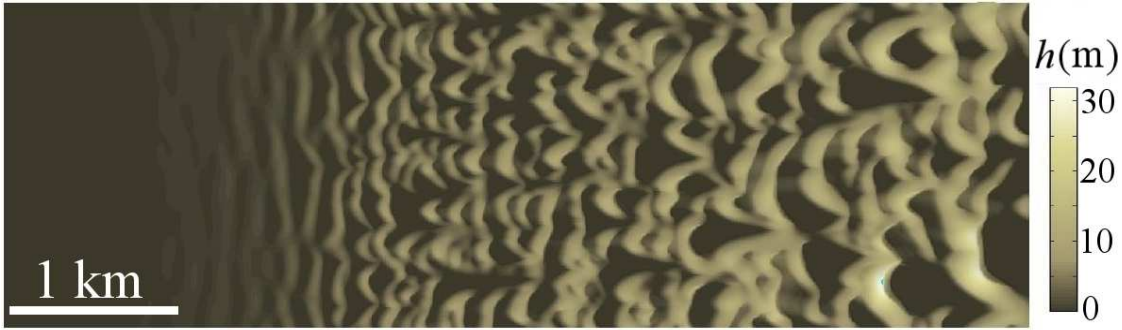}}
\caption{Snapshots of the calculation of dune field genesis, with no water table, after $\sim\!\!1000$ years. The calculation was performed with constant wind shear velocity $u_{\ast} = 0.36$~m$/$s. Wind direction is from left to right.}
\label{fig:nowater}
\end{center}
\end{figure*}

The calculations including the water table are also performed using a saturated flux condition at the inlet, i.e. $q_{\mathrm{in}} = q_{\mathrm{s}}$. Furthermore, the initial surface is adapted by including an upwind area of width ${\Delta}x$ protected from erosion, c.f. Fig.~\ref{fig:initialsurface}. This area represents a simple model for the backshore of an accreting beach \citep{Dingler_2005}, i.e. the zone immediately adjacent to the sea cliff that lies above the high-water line. At the backshore there is less potential to remove sand thus favouring accumulation \citep{Tsoar_2000}. In fact, if the water table rises above the sand upwind of the crest of the hill, as illustrated in Fig.~\ref{fig:initialsurface}, then, at a later time the sand surface is fully subdued by the water level as the hill flattens. In this case, input of sand ceases and no dune field forms. In order to avoid this problem and to make the model more realistic, we consider that the sand hill has a constant profile in the upwind backshore area, which has width ${\Delta}x \approx 25$~m. Whereas in reality the sand surface close to the beach and water table may have profiles much more complex than in the model depicted in Fig.~\ref{fig:initialsurface} \citep{Kocurek_et_al_2001}, modeling of the detailed process leading to the accumulation of sand at the backshore and the formation of the sand beach is out of the scope of the present work. Indeed, using different profiles for the sand beach or different values of ${\Delta}x$ does not change the results presented in the next Section, provided there is a sufficient amount of sand above the water level that serves as an upwind source of sand for the dune field.


\section{\label{sec:results}Results and discussion}

Next, the genesis of dunes is studied by accounting for the presence of a water table that oscillates as described in Section \ref{sec:model}. The seasonal rise and sink of the water table constrains dune dynamics. The evolution of the field depends on the quantities controlling the dynamics of the water table, namely the time period ($T_{\mathrm{w}}$) and the amplitude ($H_{\mathrm{w}}$) of the oscillation (c.f. Eq. (\ref{eq:water_table})).

One relevant quantity for dune field morphology is the time cycle, $T_{\mathrm{w}}$, of the water table relative to the migration or turnover time, $T_{\mathrm{m}}$, of the barchan, i.e. the time needed for the barchan to cover a distance approximately equal to its own width \citep{Allen_1974}. The turnover time of a barchan of height $H$ is given by the equation \citep{Hersen_et_al_2004,Duran_et_al_2010}, 
\begin{equation}
T_{\mathrm{m}} \approx a\frac{H^2}{Q},
\end{equation}
where $Q(u_{\ast}) = q_{\mathrm{s}}/{\rho}_{\mathrm{b}}$ is the bulk sand flux associated with the wind shear velocity $u_{\ast}$, and the proportionality constant, $a$, is approximately equal to $3$ \citep{Parteli_et_al_2011}. In the limit where $T_{\mathrm{w}}$ is much larger than $T_{\mathrm{m}}$, dunes are completely leveled off during inundation since their migration velocity is much faster than the oscillation rate of the water table. Sand above water level is spread throughout the field forming extensive flat regions in the wet season whereas the wind has ample time to reshape the dunes again the dry season. Conversely, if $T_{\mathrm{w}} \ll T_{\mathrm{m}}$, there is not sufficient time to build new dunes when the water level sinks below the surface. The amount of sand taken from the dunes during the wet season is small, thus leading to negligible accumulation in interdune areas. Transport between dunes in the dry season is correspondingly small, such that isolated dunes separated by large flat areas dominate the morphology of the field. 

An interesting scenario occurs when $T_{\mathrm{w}}$ becomes of the order of $T_{\mathrm{m}}$. In the wet season, the lowest portions of the dunes, i.e. the limbs, remain under water while the dune crest can migrate some distance downwind, thus leading to a decrease in dune height. Indeed, the amount of sand deposited at the interdune is large enough to construct small dunes in the dry season. Due to their fast relative migration velocity, the small bedforms emerging between the dunes collide with the larger dunes in their front, thus linking the barchans laterally and leading to the formation of barchanoidal chains. Therefore, the typical morphology observed at Len\c{c}\'ois, with chains of barchanoids alternating with interdune water ponds, arises when the size of dunes is such that their reconstitution time ($T_{\mathrm{m}}$) is of the order of the time cycle of the water table ($T_{\mathrm{w}}$). Indeed, taking $Q \approx 210$~m$^2/$year, which corresponds to $u_{\ast} = 0.36$ m$/$s, and $T_{\mathrm{w}} = 1$ year, the condition $T_{\mathrm{m}} \sim T_{\mathrm{w}}$ is fulfilled by dunes with heights of the order of $10$ m, which is consistent with the scale of dunes at Len\c{c}\'ois. 

In Fig.~\ref{fig:Tw} we present snapshots of calculations obtained with $u_{\ast} = 0.36$ m$/$s and different values of $T_{\mathrm{w}}$, namely 1 month (a), 1 year (b) and 10 years (c). Each one of Figs.~\ref{fig:Tw}a, \ref{fig:Tw}b and \ref{fig:Tw}c shows one snapshot taken at a time where the water level is maximum, i.e. $h_{\mathrm{w}} = H_{\mathrm{w}}$ (bottom), and another for the case $h_{\mathrm{w}} = -H_{\mathrm{w}}$ (top), where $H_{\mathrm{w}}$ has the nominal value 1 m. We see that the case which leads to a field of barchanoids is the one in Fig.~\ref{fig:Tw}b, i.e. when $T_{\mathrm{w}}$ has the realistic value corresponding to a seasonal oscillation.
\begin{figure*}[!ht]
\begin{center}
{\includegraphics[width=0.8 \textwidth]{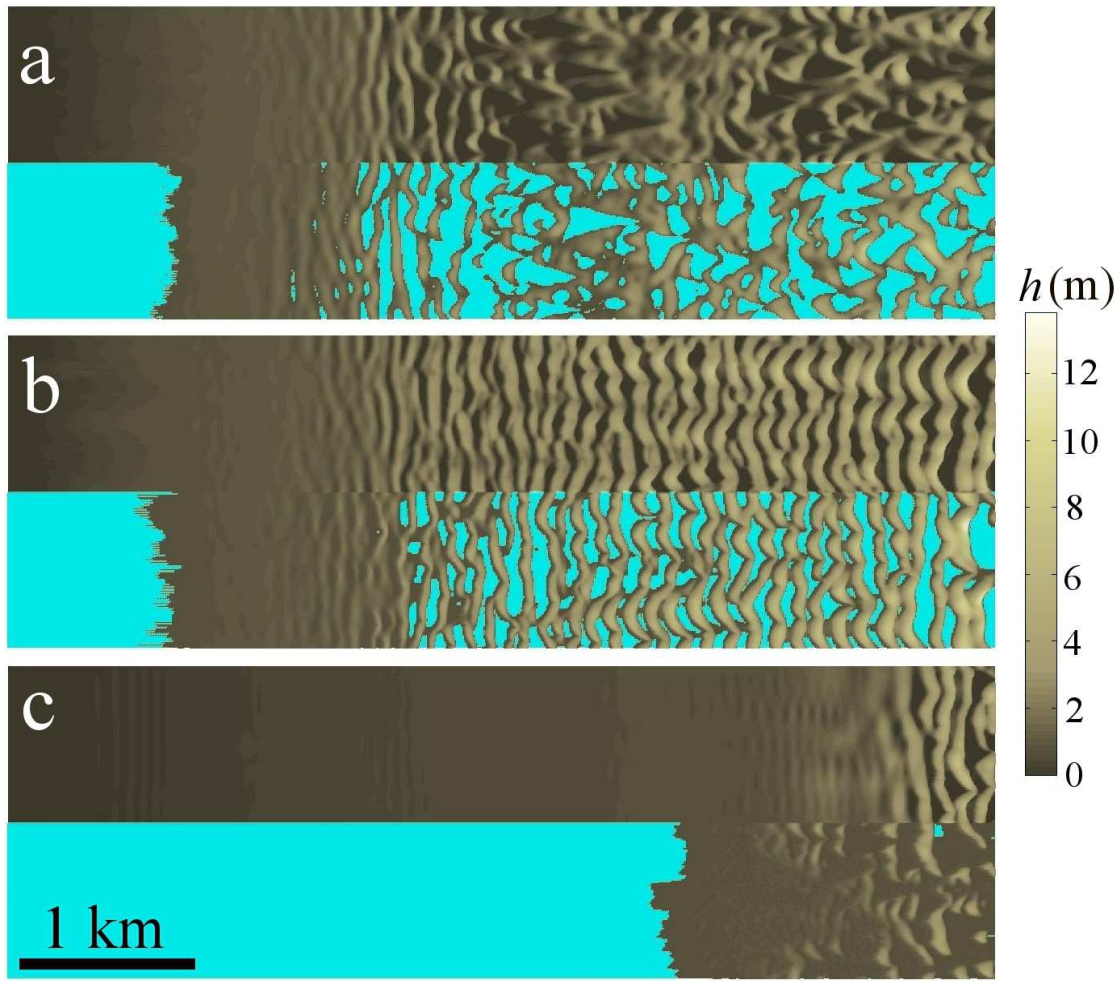}}
\caption{Snapshots of the calculation of dune field genesis after $\sim\!\!1000$ years, obtained with wind shear velocity $u_{\ast} = 0.36$~m$/$s and amplitude of the water table $H_{\mathrm{w}} = 1.0$~m. The period of the water table oscillation, $T_{\mathrm{w}}$, is (a) $0.1$ year, (b) $1.0$ year and (c) $10$ years. In each figure, the images on top and on bottom show calculation snapshots when the water level is minimum and maximum, respectively. The blue color represents the water. Wind direction is from left to right.}
\label{fig:Tw}
\end{center}
\end{figure*} 

The morphology of the field also depends on the maximum height, $H_{\mathrm{w}}$, reached by the water. Figure~\ref{fig:Hw} shows snapshots of simulations performed using the nominal value $T_{\mathrm{w}} = 1$ year and different values of $H_{\mathrm{w}}$: $20$~cm (a), 1~m (b) and 2~m (c). 
\begin{figure*}[!ht]
\begin{center}
{\includegraphics[width=0.8\textwidth]{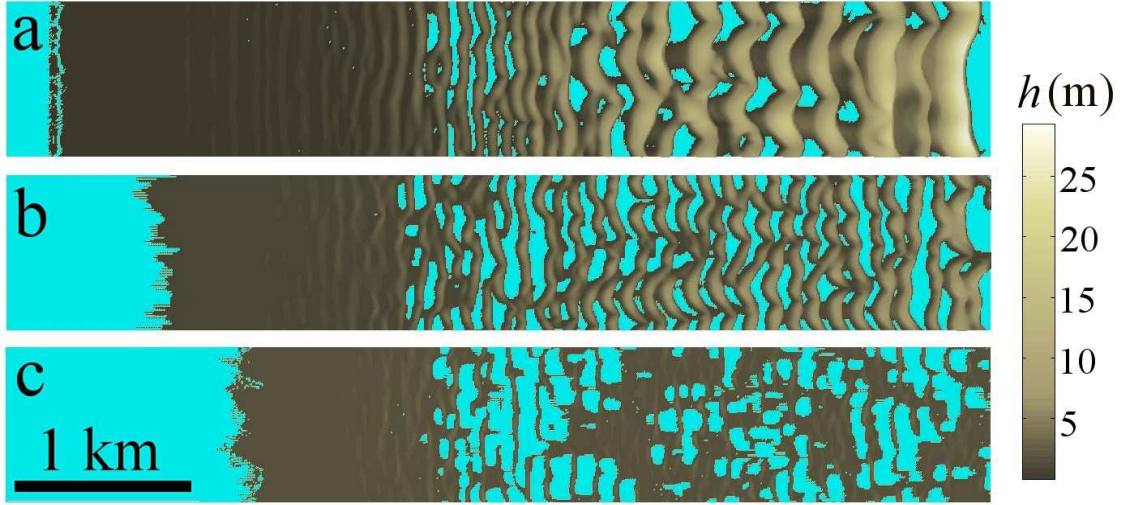}}
\caption{Snapshots of the calculation of dune field genesis after $\sim\!\!1000$ years, obtained with wind shear velocity $u_{\ast} = 0.36$~m$/$s and time cycle of the water table $T_{\mathrm{w}} = 1.0$~year. The amplitude of the water table oscillation, $H_{\mathrm{w}}$, is (a) 20~cm, (b) $1.0$~m and (c) $2.0$~m. The calculation snapshot shown in each case corresponds to an instant where the water level is maximum. The blue color represents the water. Wind direction is from left to right.}
\label{fig:Hw}
\end{center}
\end{figure*} 
As can be seen from Fig.~\ref{fig:Hw}, the average size of dunes increases with distance at a slower rate as $H_{\mathrm{w}}$ becomes larger. The average dune height at a distance of 5~km from the beach is about 28~m when $H_{\mathrm{w}} = 20$~cm, and 12~m for when $H_{\mathrm{w}} = 1$~m. For $H_{\mathrm{w}} = 2$~m, the dune height is roughly constant throughout the field and is about 4~m. Figure \ref{fig:H_vs_distance} shows the dune height at $x \approx 5$~km as a function of $H_{\mathrm{w}}$ for different values of the wind shear velocity, $u_{\ast}$. We see that $u_{\ast}$ and $H_{\mathrm{w}}$ play opposite roles for the height of dunes, i.e. while dunes become larger for increasing values of shear velocity, the water table reduces the spatial gradient of dune height in the direction of sand transport.
\begin{figure}[!ht]
\begin{center}
{\includegraphics[width=0.78 \columnwidth]{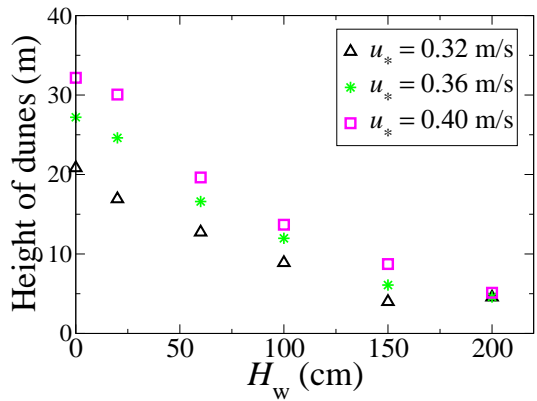}}
\caption{Dune height at a distance of $\sim\!5$~km from the beach as a function of the maximum water level, $H_{\mathrm{w}}$, for different values of the wind shear velocity, $u_{\ast}$.}
\label{fig:H_vs_distance}
\end{center}
\end{figure} 
\begin{figure}[!t]
\begin{center}
{\includegraphics[width=0.78\columnwidth]{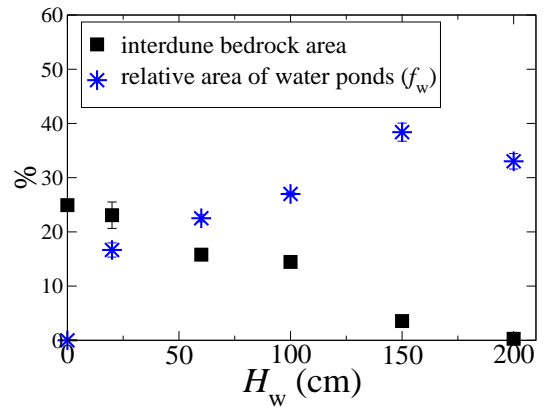}}
\caption{The plot shows the relative interdune area when the water level is minimum (squares) and the relative area of interdune water ponds when the water table reaches its maximum height (stars), for different values of the oscillation amplitude, $H_{\mathrm{w}}$. The calculations were performed with $u_{\ast} = 0.36$~m$/$s and $T_{\mathrm{w}} = 1$~year, and the results shown in the figure refer to the area between $2.8$ km and $5.6$ km downwind from the beach. The data were averaged over $10-20$ snapshots at time instants between 500 and 4000 years, when the water level was larger (stars) and smaller (square) than $95\%$ of $H_{\mathrm{w}}$ and $-H_{\mathrm{w}}$, respectively.}
\label{fig:f_w}
\end{center}
\end{figure} 

Dune height increases with distance more slowly in the presence of an oscillating water table due to cyclic phases of destruction and construction associated with the rise and sinking of the water level. When the water table rises above the ground, sand blown from the dunes accumulates into wet interdune areas, leading to flattening of the dunes. The sinking of the water table in the dry season exposes the sand deposited in the lowest areas, which serve, then, as internal source for dunes throughout the field. The maximum level, $H_{\mathrm{w}}$, reached by the water table within the wet season determines the net amount of sand blown from dunes into flatlands, and thus the evolution of dune height in the field. 

Construction and destruction of dunes due to a fluctuating water table have been documented from field observations at the back-island dune fields on Padre Island, Texas \citep{Kocurek_et_al_1992}. These fields constitute areas where the genesis and formative stages of dunes in the presence of a water table can be observed. Early portions of the fields on Padre Island are reduced to a nearly planar surface during the winter, when rainfall exceeds evaporation and the water table rises. Dunes are then reconstructed during the summer when groundwater is below the surface \citep{Kocurek_et_al_1992}. As on Padre Island, the early landscape of the dune fields produced in the simulations consists of 50 cm high bedforms, which are nearly leveled in the wet season and then reform in the dry season. As also noted by \cite{Kocurek_et_al_1992}, dunes in the ``mature'' stage of the field far downwind from the beach have more chance of survival during wet seasons due to their larger size. Dune behaviour depends thus on the maximum water level relative to the equivalent sand thickness, $\delta$, i.e. the thickness of sand if the barchan was leveled \citep{Fryberger_and_Dean_1979}. Since the volume of a barchan is given by the equation $V \approx cW^3$, with $c \approx 0.017$ \citep{Duran_et_al_2010}, and since $V \approx \delta \cdot W^2$, we obtain,
\begin{equation}
\delta \approx cW,
\end{equation}
which can be also written, using the linear relation between $W$ and the dune height, $H$ \citep{Duran_et_al_2010}, as $\delta \approx c \cdot [12H+5]$. Therefore, the smallest dunes of width $\sim 10$~m and height $\sim 1$~m are nearly leveled in the wet season when $H_{\mathrm{w}}$ is around $20-30$~cm.
 
The effect of the oscillation amplitude of the water on the morphology of the field can be further elucidated through Fig.~\ref{fig:f_w}. This figure shows that the total interdune area during the dry season decreases with the amplitude of the water table oscillation. The larger $H_{\mathrm{w}}$ the larger the amount of sand deposited in the interdune area and the smaller the relative area that is free of sand in the dry season. In contrast, the relative area of exposed water during maximum water level increases with $H_{\mathrm{w}}$, provided $H_{\mathrm{w}}$ is not too large. Figure~\ref{fig:f_w} suggests that, for intermediate values of $H_{\mathrm{w}}$ smaller than $1.5$ m, the relative interdune area can serve as a proxy for the maximum water level relative to dune height at a given location of a dune field. When the value of $H_{\mathrm{w}}$ increases beyond $1.5$~m, the fraction of dune surface above the water level becomes so small that erosion of dune crest leads to flat sand sheets of thickness comparable to the dune height. As shown in Fig.~\ref{fig:Hw}c, these interdune sand sheets can fill large areas of the lagoons between dunes, thus leading to a decrease in the value of $f_{\mathrm{w}}$ as the water level becomes larger. Indeed, the average value of $f_{\mathrm{w}}$ for $H_{\mathrm{w}} = 2.0$~m is slightly smaller than the one for $H_{\mathrm{w}} = 1.5$~m, as shown in Fig.~\ref{fig:f_w}. For such large values of $H_{\mathrm{w}}$, the morphology of the field deviates from the typical landscape of barchanoidal chains observed at Len\c{c}\'ois Maranhenses (c.f. Fig.~\ref{fig:Hw}). Therefore, in order to make a quantitative comparison of the simulation results with dunes at Len\c{c}\'ois, we take smaller values of $H_{\mathrm{w}}$, namely around 1~m. 

Figure~\ref{fig:comparison_Lencois_1}a shows an image of barchanoidal dunes at Len\c{c}\'ois Maranhenses, which are located within area I in Fig. 1a, downwind of a recently investigated field of transverse dunes of heights between 7~m and 10~m \citep{Parteli_et_al_2006}. In Fig.~\ref{fig:comparison_Lencois_1}b, we show a snapshot of a calculation obtained with $H_{\mathrm{w}} = 60$ cm, while $u_{\ast} = 0.36$ m$/$s and $T_{\mathrm{w}} = 1$ year are taken in consistence with the conditions at Len\c{c}\'ois Maranhenses. We see that good quantitative agreement is found between dunes obtained in the simulations and real dunes. In both Figs.~\ref{fig:comparison_Lencois_1}a and \ref{fig:comparison_Lencois_1}b we count 14 dune crests along the longitudinal cut indicated in the corresponding images. Furthermore, we calculate the total relative area of the terrain ($f_{\mathrm{w}}$) which is covered by water, and find $f_{\mathrm{w}} = 28\%$ and $31\%$ for the dunes in the image and in the simulation, respectively.
\begin{figure}[!ht]
\begin{center}
{\includegraphics[width=0.74 \columnwidth]{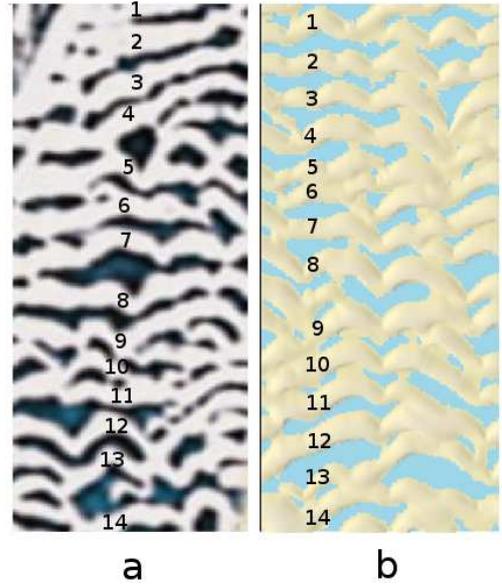}}
\caption{(a) Image of barchanoids at Len\c{c}\'{o}is Maranhenses, within an area of 986~m $\times$ 1964~m; (b) barchanoids produced in the simulation using $u_{\ast} = 0.36$ m$/$s, $H_{\mathrm{w}} = 100$~cm and $T_{\mathrm{w}} = 1.0$~year.}
\label{fig:comparison_Lencois_1}
\end{center}
\end{figure}

Further downwind in the field of Len\c{c}\'ois Maranhenses, in area II of Fig. 1a, dunes are larger and can reach heights of about $15-20$~m. In Fig.~\ref{fig:comparison_Lencois_2}a, an image of barchanoids within area II is shown, whereas a snapshot of a simulation performed with $H_{\mathrm{w}} = 1$ m is displayed in Fig.~\ref{fig:comparison_Lencois_2}b. Again, the calculation outcome is in good quantitative agreement with the image of the real dune field. The value of $f_{\mathrm{w}}$ calculated for the real dunes is about $34\%$, while for the dunes in the simulation we obtain $f_{\mathrm{w}} = 28\%$. 
\begin{figure}[!ht]
\begin{center}
{\includegraphics[width=0.74 \columnwidth]{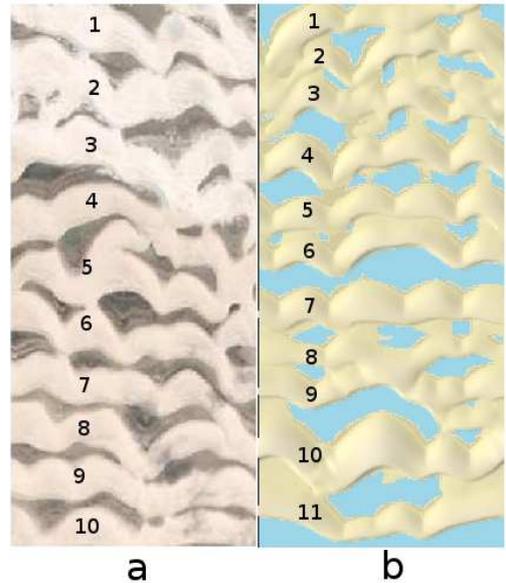}}
\caption{(a) Image of barchanoids at Len\c{c}\'{o}is Maranhenses, within an area of 986~m $\times$ 1964~m; (b) barchanoids produced in the simulation using $u_{\ast} = 0.4$ m$/$s, $H_{\mathrm{w}} = 60$~cm and $T_{\mathrm{w}} = 1.0$~year.}
\label{fig:comparison_Lencois_2}
\end{center}
\end{figure} 

Therefore, the relative surface area ($f_{\mathrm{w}}$) that is covered by water in the wet season as obtained from our calculations is about $30\%$, which is close but below the value of $41\%$ corresponding to the total area of Len\c{c}\'ois \citep{Levin_et_al_2007}. However, it has to be emphasized that the value of $f_{\mathrm{w}}$ estimated from our calculations corresponds only to interdune ponds between chains of barchanoids (c.f. Figs.~\ref{fig:comparison_Lencois_1} and \ref{fig:comparison_Lencois_2}). In reality, the area of Len\c{c}\'ois also hosts large permanent lagoons, indeed in some portions of the field the sand is distributed in an irregular manner leading to interdune flats of different sizes, which further contributes to the formation of well-separated barchans and a larger value of $f_{\mathrm{w}}$. 

We have also considered an oscillating wind shear velocity that is negatively correlated with the water table, i.e. $u_{\ast}(t) = {\bar{u}}_{\ast} - U_{\ast}{\mbox{sin}}[2{\pi}t/T_{\mathrm{w}}]$, with $T_{\mathrm{w}} = 1$ year, ${\bar{u}}_{\ast} \approx 0.30$ m$/$s and $U_{\ast} \approx 0.1$ m$/$s \citep{Jimenez_et_al_1999,Tsoar_et_al_2009}. However, calculations using this model for wind velocity oscillating out of phase with the water table led to dune patterns similar to those obtained from simulations with a constant wind strength. Therefore, we conclude that the oscillation in the water table is the relevant factor controlling the morphology of dunes at Len\c{c}\'ois, rather than the seasonal modulation of the wind strength. 

%
%

\section{\label{sec:conclusions}Conclusions}

We extended the dune model introduced by \cite{Sauermann_et_al_2001} to study dune formation in presence of a seasonally varying water table. The model was applied to investigate the genesis of dunes from a sand source in an environment where groundwater level oscillates in response to variations of rainfall and evapotranspiration rates. We found that the morphology of the field is dictated by the quantities controlling the dynamics of the water table, namely the time period ($T_{\mathrm{w}}$) and the amplitude ($H_{\mathrm{w}}$) of the oscillation. Chains of barchanoids, such as those at Len\c{c}\'ois Maranhenses dune field in northeastern Brazil, are obtained when $T_{\mathrm{w}}$ is of the order of the dune turnover time, while $H_{\mathrm{w}}$ determines the rate at which dune height increases in wind direction. We could reproduce barchanoidal dunes with the same size as those at Len\c{c}\'ois by using realistic values for $T_{\mathrm{w}}$ (1 year) and $H_{\mathrm{w}}$ (between 60 cm and 1 m) and an average wind shear velocity $u_{\ast} = 0.36$ m$/$s that is within the ranges observed at the real field. Furthermore, we found that the maximum water level affects the total relative interdune area in the dune field, $f_{\mathrm{w}}$. Our calculations yield an estimate for $f_{\mathrm{w}}$ at Len\c{c}\'ois which is close to the value reported from measurements.

Several factors should be included in the future in order to improve the quantitative assessment of dunes at Len\c{c}\'ois Maranhenses. While the calculations of the present work considered a constant (saturated) sand influx, an improved calculation of the genesis and formation of Len\c{c}\'ois should consider a sand influx which may vary in time and also along the transverse profile of the beach. In some portions of the field, transport rates may change significantly due to the occurence of permanent lagoons or areas with quicksand \citep{Kadau_et_al_2009}, leading to spatial changes in sand availability. Variations of local topography may also contribute to the changes in sediment budget, dune height and local morphology of the field. Moreover, the present model must be improved in the future in order to account for the capillary fringe of the water table, i.e. the wet layer of sand immediately above the water table. As a matter of fact, when the water table is below the sand surface, groundwater seeps up by capillary action to fill the pores, such that a subsurface layer (the capillary fringe) becomes saturated with water and thus protected from erosion. By enhancing moisture concentration and sediment cohesion, the capillary fringe may alter local rates of erosion and deposition throughout the field and thus influence the shape of dunes and interdune areas \citep{Schenk_and_Fryberger_1988,Grotzinger_et_al_2005}. Finally, it would be interesting to combine the present model for dune formation in presence of a water table with the model for sand transport with vegetation growth \citep{Duran_and_Herrmann_2006a,Duran_et_al_2008,Luna_et_al_2011} in order to investigate the combined effect of those natural agents on the genesis and evolution of coastal dune fields. 

\section*{Acknowledgments}
This work was supported in part by Brazilian agencies FUNCAP, CAPES, FINEP and CNPq, by the CNPq/FUNCAP Pronex Grant, by Swiss National Foundation Grant NF 20021-116050/1 and ETH Grant ETH-10 09-2.





\end{document}